# Out-of-contact elastohydrodynamic deformation due to lubrication forces

Yumo Wang[1], Charles Dhong[1], and Joelle Frechette[1,2]

[1]Chemical and Biomolecular Engineering Department, and [2]Hopkins Extreme Materials Institute

Johns Hopkins University, Baltimore MD 21218.

(Dated November 1, 2015)

**Abstract**

We characterize the spatiotemporal deformation of an elastic film during the radial drainage of fluid from a narrowing gap. Elastic deformation of the film takes the form of a dimple and prevents full contact to be reached. With thinner elastic film the stress becomes increasingly supported by the underlying rigid substrate and the dimple formation is suppressed, which allows the surfaces to reach full contact. We highlight the lag due to viscoelasticity on the surface profiles, and that for a given fluid film thickness deformation leads to stronger hydrodynamic forces than for rigid surfaces.

Surface and interfacial phenomena in soft matter display complex mesoscale behaviors that are qualitatively different from those encountered in stiff materials, such as elastic instabilities during adhesion[1,2] and Schallamach waves in friction[3,4]. Surface [5-7] or viscous[8] stresses can also lead to elastic deformations that are similar to those observed at fluid interfaces. Elastohydrodynamic deformation (EHD), for example, can cause lift and reduce friction during sliding [9-13] and alter the rheological properties of soft colloidal particles[14-17]. Elastohydrodynamic deformation also modifies the shape of approaching surfaces, a determining factor for the adhesion dynamics to wet or flooded surfaces.[18-21] When studying elastohydrodynamics in soft matter it is a challenge to measure simultaneously the hydrodynamic forces and the deformation, both necessary to understand how contact is reached and the coupling between deformation and viscous dissipation.

To illustrate the importance of elastohydrodynamic deformations, consider the normal approach of a rigid sphere toward a surface with an elastically compliant coating in a Newtonian fluid (Fig. 1A). The hydrodynamic forces lead to deformation of the soft material prior to contact ($w(r,t)$), as was visualized by Roberts during the settling of a rubber sphere toward a wall.[8] For elastic half-space this problem can be described by the theory of Davis et al.[22,23] derived for the collision of elastic spheres in fluid, and based on the coupling between lubrication forces and linear elasticity. Recent direct measurements of viscous forces in the presence of a soft surface demonstrated that even minute elastic deformations can have a profound effect on the hydrodynamic interactions.[24,25] Therefore, elasticity likely has to be considered when studying slip at a solid-liquid interfaces. The predominance of soft coatings in tribology and adhesion makes the extension of elastohydrodynamic theory to thin supported films technologically relevant, especially to understand how contact is reached in soft matter. The treatment for supported elastic films, however, is challenging and has limited experimental validation. For thin films (thickness $<< \sqrt{2Rh}$ ), the underlying substrate can support a significant fraction of the mechanical stress, which can alter the elastohydrodynamic response from that expected with semi-infinite solids[26,27]. The theory for supported films developed by Charlaix, for instance, elegantly takes advantage of the contribution of the underlying substrate on the hydrodynamic forces to extract the Young's modulus of coatings.[25,27,28] However, the absence of absolute measurement of spatiotemporal separation brings uncertainties to the role played by elasticity on hydrodynamic interactions, especially for the case of thin elastic coatings where our understanding is more limited. Combining visualization of spatiotemporal deformation with force measurements would allow to understand the dynamic of contact formation in soft materials, and to analyze the response of supported films.

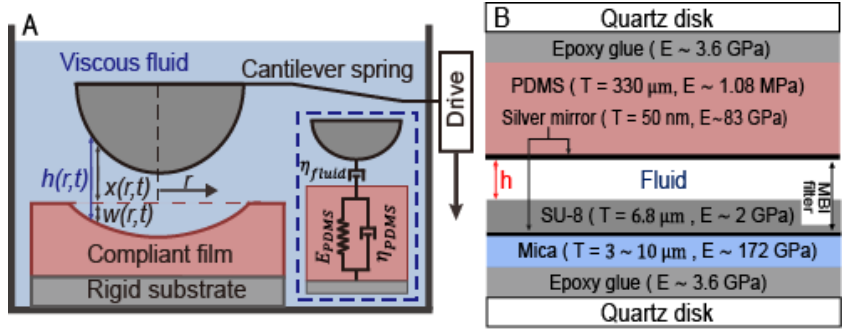

FIG. 1. Schematics (not to scale) of (A) the elastohydrodynamic problem with labelled variables (Inset: Kelvin-Voigt model for elastomer viscoelasticity), and (B) Material layers and properties.

In this letter we investigate the role of compliance on the normal approach of a soft surface toward a rigid one in a viscous Newtonian fluid (Fig. 1A). Spatiotemporal deformation profiles and hydrodynamic forces are measured, and compared to an elastohydrodynamic theory for half-space. We find that elastic deformation in the shape of a dimple at the centerpoint prevents contact between approaching surfaces. We also observe that the finite thickness of the elastic layer restricts the deformation and favors contact. Finally, we show that deformation leads to significantly stronger hydrodynamic forces than those observed with rigid surfaces for the same central separation.

Experiments are performed between crossed-cylinders (equivalent to the sphere-plane geometry) using the Surface Forces Apparatus.[29-31] One surface is rigid (bottom in Fig. 1B) and the other is compliant due to the presence of a relatively thick 330 μm PDMS film (polydimethyl siloxane) coated with a 50 nm silver film as a top layer (top in Fig. 1B). Both surfaces are glued on a cylindrical disk (radius, R=1.75 cm). The top silver film facilitates interferometry and prevents swelling in the silicone oil (viscosity, η=0.2Pa*s). An effective Young's modulus of 1.08 ± 0.05 MPa for the PDMS film was obtained by performing *in situ* contact mechanics experiments[32-34] in silicone oil with the same surfaces (see supporting information 2[35-40]). Because of the underlying rigid substrate [41-45], we expect this modulus to overestimate the intrinsic modulus of the PDMS layer by 15-20%.[46,47] We rely on white light multiple-beam interferometry[30,48,49] to map the local fluid film thickness, $h(r,t)$, within nanometer resolution in the normal direction and micron resolution in the lateral direction.

The dynamic experiments follow the approach of Chan and Horn[50,51], where a disk initially at rest and mounted on a cantilever spring (spring constant k = 165.3 N/m) is driven toward the other surface at a constant drive velocity ($V$). The spring deflects because of the drag, and the velocity of the surface ($v$) is always less than the drive velocity. As the surfaces approach, the hydrodynamic forces increase and deform the PDMS film, as evidenced by the flattening at the center, see III-IV in Fig. 2A. Further approach lead to an increase in the fluid pressure near the center causing the formation of a dimple in the elastic film, see V-VII in Fig. 2A.

For the theoretical description we employ the lubrication equation in axisymmetric coordinates ($h \ll \sqrt{2Rh}$) and follow closely the treatment of Ref [22] to couple the fluid pressure distribution ($p(r,t)$) with linear elasticity of the compliant film. We treat the elastic film as a half-space in the small strain limit (strain of the PDMS coating here, $\varepsilon < 0.5\%$), i.e. we neglect the contribution of the substrate supporting the elastic film. We incorporate a force balance, $F(t) = k[h(0,t) - Vt - h(0,0) - w(0,t)] = \int_0^R 2\pi p(r,t) r dr$, where the cantilever spring deflects due to the repulsive viscous forces, $F(t)$. Here $h(0,0)$ is the initial separation at the centerpoint. We neglect the radial shear stress on the film and use the no-slip boundary condition for both surfaces. We obtain a solution numerically using the initial fluid film profile ($h(r,0)$) from the experiments as the initial condition without any fitting parameters. As a second description we treat the PDMS film as a viscoelastic material with a viscosity $\eta_{PDMS}$, and model the film's response to an applied load as a spring and dashpot in parallel (Kelvin-Voigt model, Fig. 1A). In the viscoelastic description $\eta_{PDMS}$ is not known *a priori* and we iterate to find a single $\eta_{PDMS}$ that best describes all the profiles for all drive velocities. (see supporting information [35-40] for details of the model, algorithm, and treatment of viscoelasticity.)

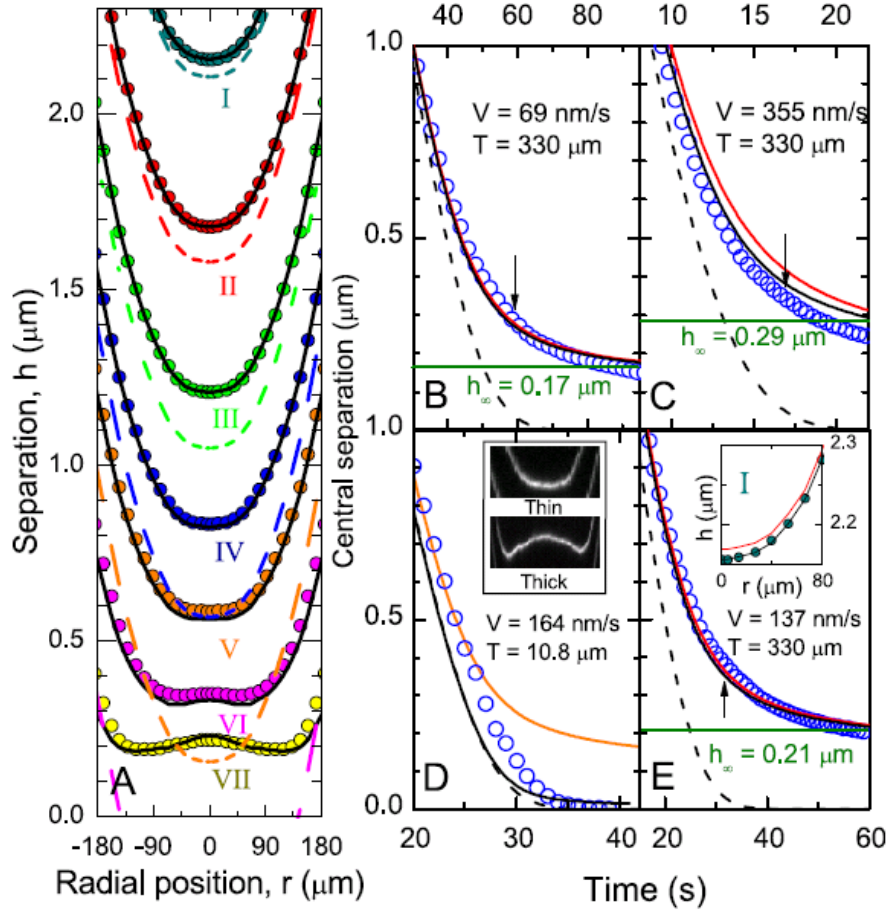

FIG. 2. (A) Experimental and theoretical spatiotemporal surface profile during approach at V = 137 nm/s. The black solid lines correspond to theoretical predictions treating the PDMS films as a viscoelastic solid. Time stamps are: I: t = 3.8s, II: t = 8.8s, III: t = 13.8s, IV: t = 18.8s, V: t = 23.8s, VI: t = 33.8s, and VII: t = 53.8s. Dash lines are for the positions of the corresponding undeformed sphere. (B-E): Temporal central separation for: (B) V = 69 nm/s, (C) V = 355 nm/s, (D) 164 nm/s, and (E) V = 137 nm/s. (B-C): Effect of drive velocity. (D-E): Effect of film thickness. Black solid lines are the same as in (A), dash lines: Reynolds' theory. Red solid lines are predictions treating the elastomer as an elastic solid. Black arrows: time for dimple formation. $h_\infty$ : Long time predictions (central dh/dt < 1%V). (D): Approach of a thinner PDMS coating (T = 10.9 μm, R = 1.10cm), black rigid line represents predictions for E = 84 MPa. Yellow line represents the predictions for E = 1 MPa. Insets of: (D) shape of fringes for thin (10.9 μm) and thick (330 μm ) PDMS film during the approach with $h_{center}$ = 150 nm and (E) Effect of viscosity of PDMS on initial surface profile.

The measured and predicted profiles are shown in Fig. 2. In general the elastic solution is sufficient to describe the surface profile but treating the PDMS as a viscoelastic solid gives a better agreement. The viscoelasticity of the PDMS alters the fluid film profile when the rate of strain is the largest (acceleration and deceleration) such as during start up where viscoelastic contributions are visible (inset of Fig. 2E and supporting information [35-40]). For the viscoelastic predictions, a single value of $\eta_{PDMS} = 1.5 MPa \cdot s$ best fits all the profiles at all velocities, in agreement with literature values.[52] In Fig. 2A, the predictions with viscoelasticity predict fluid film thicknesses that are always ±35nm of the measured values at the centerpoint. The error increases with drive velocity: at 355nm/s it is ±48nm, while it is less than 30nm for 69nm/s. For all drive velocities when the two surfaces are close (strong hydrodynamic forces), the observed separation is less than predicted. This error can be understood considering that surfaces appear stiffer as the forces increase due to the finite thickness of the elastomer, and at a constant time stiffer surfaces are always closer than compliant ones (inset of Fig 4).

Elastic deformation prevents the surfaces from reaching contact at all drive velocities investigated, which is captured by the long time predictions (central dh/dt < 1%V, see Fig. 2B,C,E). As the surfaces approach, flattening away from the centerpoint occurs faster than the normal motion toward the surface, which leads to dramatically large forces and prevents contact. Theoretical solutions for the surface separation are not defined at contact regardless of compliance. For rigid materials, predictions diverge at very short-range where irreversibilities such as roughness, size of fluid molecules, and surface forces often favor contact in experiments[53]. In contrast, for a compliant material, the separation at long times is sufficiently large to prevent these mechanisms from playing a role. With compliant surfaces the drive will lead to a broader surface instead of significantly decreasing the central fluid film thickness, at least until non-linear effects occur. Note that contact can be reached under quasi-static condition.

The thickness of the compliant layer plays an important role in determining the spatiotemporal fluid film thickness. We contrast the temporal change in surface separation at the centerpoint of a thick (T=330 μm, Fig. 2E) and thin (T=10.9 μm, Fig. 2D) PDMS films for similar drive velocities. Both films have the same bulk mechanical properties, however the effective modulus is much larger (E=84 MPa) for the thin film because of incompressibility and apparent stiffening due to the underlying rigid substrate (supporting information[35-40])[54,55]. For the thin film, as the hydrodynamic forces increase, the stress becomes increasingly supported by the rigid substrate. As a result, the temporal fluid film thickness gradually transitions from being the one predicted for a compliant material (E=1MPa) to that of a rigid one (see predictions for the two moduli in Fig. 2D). We find that the effective stiffening suppresses the formation of a dimple (within our spatial resolution) in the elastic film, and that contact can be reached in a fashion similar than for rigid materials. Such a transition to a rigid-like behavior is not observed with the thicker film. This stiffening effect is well-characterized for contact mechanics experiments[45]. Our work shows how the finite thickness of the elastic film gradually alters the deformation profile from that of a semi-infinite compliant material as the surfaces approach and how it favors contact. Increasing the modulus in the model will not give better agreement with experiments, and always make the far-field predictions significantly worse (see supporting information[35-40]). An alternative treatment would be to use a solution for arbitrary axisymmetric pressure distribution for a finite thickness elastic layer, such as in Refs [26,27], to obtain a solution valid at all $h$. A simplified scaling argument treating the deformation solely as shear, such as in Ref [27] could also work close to contact but not for the far-field. The importance of film thickness on the force required to make contact has profound implications for hydrodynamic interactions with soft materials and coatings, such as in biological systems, tribology, adhesion, and rheology.

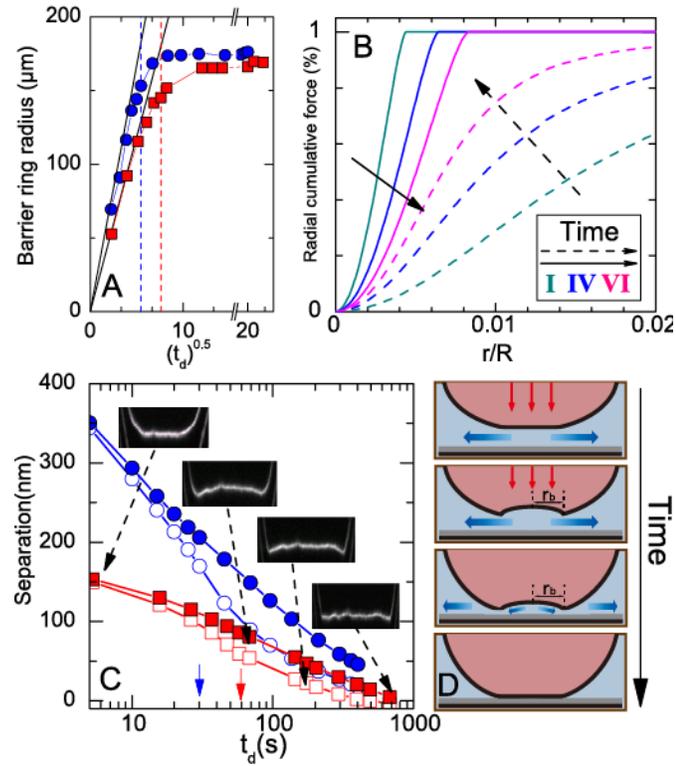

FIG. 3. (A) Growth of barrier ring radius ($r_b$). Squares: V = 69 nm/s, circles: V = 137 nm/s. $t_d$ (s) is the time elapsed after center curvature of the elastomer becomes negative. Black solid lines: $r_b = \sqrt{RV\Delta t/2}$. Vertical dashed lines indicate when the motor stopped. (B) Radial cumulative force (%) as a function of r/R for V = 137 nm/s. The roman numerals represent the same times as those of Fig. 2A. Solid lines correspond to the relative cumulative force results from a spherical indenter with the same load as that in EHD (Dashed lines), calculated from Hertz contact mechanics. (C) Centerpoint (solid) and edge (open) separation after dimple formation (circles: V=137 nm/s, squares: V=69 nm/s). Inset: Corresponding interference fringes for V = 69 nm/s. Solid arrows: motor stop time for V = 137 nm/s and V = 69 nm/s. (D) Schematic showing formation and relaxation of dimples with a barrier ring $r_b$.

The formation of dimple—a bell of liquid trapped around the centerpoint— is observed as the force increases (Fig. 2A). Once formed ($t_d$=0) the growth of dimples forming a barrier ring $r_b$ follows the same geometric scaling as the one observed for fluid droplets, and is independent of materials properties ($r_b = \sqrt{RV\Delta t/2}$, Fig. 3A)[56,57]. This scaling implies that beyond $t_d$ the fluid film thickness remains essentially constant while the increase in pressure is almost solely accommodated by elastic deformation. The appearance of a dimple requires the fluid pressure to be highly concentrated near the centerpoint and our model (Fig. 3B) shows that as the force increases, the fluid pressure distribution becomes increasingly more concentrated near the center. We compare the radial cumulative force with the one predicted based on a Hertzian contact for the same force (Fig. 3B). For a given force, a spherical indenter always leads to a narrower pressure distribution than the elastohydrodynamic case. As the force increases, however, the contact area based on indentation increases while the elastrohydrodynamic pressure distribution becomes shaper and significantly more concentrated near the centerpoint (compare the radial cumulative force at the Hertz contact radius for the three cases shown in Fig. 3B).

If we stop the motor (near the limit of the range of the motor), the surface velocity decreases but does not stop because of the stored energy in the cantilever. The dimple slowly relaxes after the motor stops (see Fig. 3C), and after a long time contact can be reached first at the edge of the ring, followed later by near contact (to within 10nm) at the centerpoint (Fig. 3C). This process is very slow (»100s): the fluid has to drain through the edge of the dimple as the pressure drop between the center and the surrounding decreases. During this relaxation a fluid pocket can be trapped at the center while contact is reached at the edge.

The measured hydrodynamic forces and predictions for soft and rigid surfaces are shown in Fig. 4. The experimental points are calculated based on the measured fluid film profile and predictions for the model treat the PDMS film as a viscoelastic solid. To calculate the hydrodynamic force from our experimental data we used the prediction for $w$ at the centerpoint (see supporting information [35-40]). In general our experiments show excellent agreement with predictions over all the velocities, with the largest error present for the fastest drive velocity and close to contact. When comparing the hydrodynamic forces between soft and rigid surfaces we see that predictions based on rigid surfaces underestimate the real force for all fluid film thicknesses. In contrast, Reynolds theory always overestimates the force at a given time (inset of Fig. 4). For a given fluid film thickness the deformed surface is flatter, giving rise to larger hydrodynamic repulsion than for rigid surfaces. In contrast, at a given time $t$ rigid surfaces are always closer to contact and the force is higher than for a deformable surface. We also observe systematic deviation in the hydrodynamic forces at small $h$ and at long times that are attributed to the effective stiffening caused by the rigid underlying substrate. If we compare with AFM experiments where only $F(t)$ and $x(r,t)$ are known, predictions based on Reynolds theory would overestimate the measured force: for the same $x(r,t)$ a rigid surface has a smaller separation ($h$) than the compliant surface. Thus the rigid case predicts a larger force than measured because of the different $h$, as shown by Ref [24].

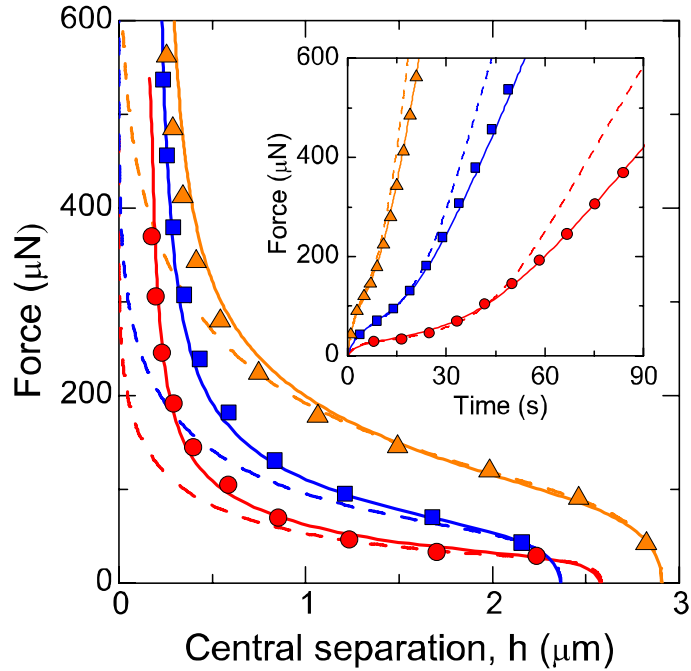

FIG. 4. Repulsive elastohydrodynamic force as a function of central separation, $h$. Circles: $V$ = 69 nm/s, squares: $V$ = 137 nm/s, triangles: $V$ = 355 nm/s. Dash lines: predictions for rigid surfaces, solid lines: predictions for compliant surfaces treating the elastomer as a viscoelastic solid. Inset: corresponding force as a function of time.

In summary, we characterized the spatiotemporal deformation of a compliant film during the normal drainage of fluid from a narrowing gap. For a thick elastic film (approx. half-space) we observe that elastic deformation in the form of a dimple prevents the surfaces from reaching contact. For a thinner elastic the formation of the dimple is suppressed and contact can be reached because the stress is supported by the underlying substrate. We find that the growth of the dimples in the elastic films is nearly independent of the mechanical properties of the film. Finally we find that at a given time elastic compliance leads to weaker forces while it leads to stronger forces at a given fluid film thickness. Measuring absolute surface separation is critical when working with soft materials, such as in biological systems or in the lubrication of surfaces with compliant coatings of a finite thickness.

**Acknowledgements.** This work is partially supported by the Office of Naval Research – Young Investigator Award (N000141110629), by the Donors of the American Chemical Society Petroleum Research Fund under Grant 51803-ND5, the Hopkins Extreme Materials Institute (HEMI), and NSF-CMMI

1538003.

**References.**

[1]  A. Ghatak and M. K. Chaudhury, J. Adhes. **83**, 679 (2007).
[2]  S. Yang, K. Khare, and P. C. Lin, Adv. Funct. Mater. **20**, 2550 (2010).
[3]  C. J. Rand and A. J. Crosby, J. Appl. Phys. **106**, 4, 064913 (2009).
[4]  K. Vorvolakos and M. K. Chaudhury, Langmuir **19**, 6778 (2003).
[5]  J. S. Wexler, T. M. Heard, and H. A. Stone, Phys. Rev. Lett. **112**, 5, 066102 (2014).
[6]  L. A. Lubbers, J. H. Weijs, L. Botto, S. Das, B. Andreotti, and J. H. Snoeijer, J. Fluid Mech. **747**, 12, R1 (2014).
[7]  R. W. Style, R. Boltyanskiy, Y. L. Che, J. S. Wettlaufer, L. A. Wilen, and E. R. Dufresne, Phys. Rev. Lett. **110**, 5, 066103 (2013).
[8]  A. Roberts, J. Phys. D: Appl. Phys. **4**, 423 (1971).
[9]  D. Dowson, Wear **190**, 125 (1995).
[10] A. Gopinath and L. Mahadevan, in *Proceedings of the Royal Society of London A: Mathematical, Physical and Engineering Sciences* (The Royal Society, 2011), p. rspa20100228.
[11] J. M. Skotheim and L. Mahadevan, Phys. Fluids **17**, 092101 (2005).
[12] M. Scaraggi, G. Carbone, B. N. J. Persson, and D. Dini, Soft Matter **7**, 10395 (2011).
[13] G. A. Ateshian, H. Wang, and W. M. Lai, J. Tribol. **120**, 241 (1998).
[14] J. R. Seth, L. Mohan, C. Locatelli-Champagne, M. Cloitre, and R. T. Bonnecaze, Nat. Mater. **10**, 838 (2011).
[15] A. Ikeda, L. Berthier, and P. Sollich, Phys. Rev. Lett. **109**, 018301 (2012).
[16] M. Gross, T. Kruger, and F. Varnik, Soft Matter **10**, 4360 (2014).
[17] S. Adams, W. Frith, and J. Stokes, J. Rheol. (1978-present) **48**, 1195 (2004).
[18] J. Iturri, L. Xue, M. Kappl, L. García‐Fernández, W. J. P. Barnes, H. J. Butt, and A. del Campo, Adv. Funct. Mater. (2015).
[19] B. N. J. Persson, J. Phys. Condens. Matter **19**, 16, 376110 (2007).
[20] L. Heepe and S. N. Gorb, Annu. Rev. Mater. Res. **44**, 173 (2014).
[21] C. Dhong and J. Frechette, Soft Matter **11**, 1901 (2015).
[22] R. H. Davis, J.-M. Serayssol, and E. Hinch, J. Fluid Mech. **163**, 479 (1986).
[23] G. Barnocky and R. H. Davis, Phys. Fluids **31**, 1324 (1988).
[24] F. Kaveh, J. Ally, M. Kappl, and H. J. Butt, Langmuir **30**, 11619 (2014).
[25] S. Leroy, A. Steinberger, C. Cottin-Bizonne, F. Restagno, L. Leger, and E. Charlaix, Phys. Rev. Lett. **108**, 264501 (2012).
[26] X. Zhao and R. Rajapakse, Int. J. Eng. Sci. **47**, 1433 (2009).
[27] S. Leroy and E. Charlaix, J. Fluid Mech. **674**, 389 (2011).
[28] R. Villey, E. Martinot, C. Cottin-Bizonne, M. Phaner-Goutorbe, L. Léger, F. Restagno, and E. Charlaix, Phys. Rev. Lett. **111**, 215701 (2013).
[29] D. Tabor and R. H. S. Winterton, Proc. R. Soc. London, Ser: A **312**, 435 (1969).
[30] J. N. Israelachvili, J. Colloid Interface Sci. **44**, 259 (1973).
[31] J. N. Israelachvili *et al.*, Rep. Prog. Phys. **73**, 036601 (2010).
[32] K. L. Johnson, K. Kendall, and A. D. Roberts, Proc. R. Soc. London, Ser. A **324**, 301 (1971).
[33] M. K. Chaudhury and G. M. Whitesides, Science **255**, 1230 (1992).
[34] K. Khanafer, A. Duprey, M. Schlicht, and R. Berguer, Biomed. Microdevices **11**, 503 (2009).
[35]  For PDMS preparation, calculation of the effective modulus and layering effect, determination of the deformation and numerical algorithm, see Supplemental Material, which include Refs.[36-40].
[36] J. N. Lee, C. Park, and G. M. Whitesides, Anal. Chem. **75**, 6544 (2003).
[37] M. Heuberger, Rev. Sci. Instrum. **72**, 1700 (2001).

[38]	G. Lian, M. J. Adams, and C. Thornton, J. Fluid Mech. **311**, 141 (1996).
[39]	J.-M. Sérayssol, MS thesis, University of Colorado, Boulder (1985).
[40]	B. Hughes and L. White, Q. J. Mech. Appl. Math. **32**, 445 (1979).
[41]	I. Sridhar, K. L. Johnson, and N. A. Fleck, J. Phys. D: Appl. Phys. **30**, 1710 (1997).
[42]	K. L. Johnson and I. Sridhar, J. Phys. D: Appl. Phys. **34**, 683 (2001).
[43]	P. M. McGuiggan, J. S. Wallace, D. T. Smith, I. Sridhar, Z. W. Zheng, and K. L. Johnson, J. Phys. D: Appl. Phys. **40**, 5984 (2007).
[44]	F. K. Yang, W. Zhang, Y. G. Han, S. Yoffe, Y. C. Cho, and B. X. Zhao, Langmuir **28**, 9562 (2012).
[45]	E. Barthel and A. Perriot, J. Phys. D: Appl. Phys. **40**, 1059 (2007).
[46]	K. R. Shull, Mater. Sci. Eng. R-Rep. **36**, 1 (2002).
[47]	E. K. Dimitriadis, F. Horkay, J. Maresca, B. Kachar, and R. S. Chadwick, Biophys. J. **82**, 2798 (2002).
[48]	S. Tolansky, *Multiple-Beam Interferometry of surfaces and films* (Oxford University Press, London, 1948).
[49]	R. Gupta and J. Fréchette, J. Colloid Interface Sci. **412**, 82 (2013).
[50]	D. Y. C. Chan and R. G. Horn, J. Chem. Phys. **10**, 5311 (1985).
[51]	R. Gupta and J. Frechette, Langmuir **28**, 14703 (2012).
[52]	S. T. Choi, S. J. Jeong, and Y. Y. Earmme, Scr. Mater. **58**, 199 (2008).
[53]	R. Horn and J. Israelachvili, Macromolecules **21**, 2836 (1988).
[54]	A. Perriot and E. Barthel, J. Mater. Res. **19**, 600 (2004).
[55]	E. Barthel, A. Perriot, A. Chateauminois, and C. Fretigny, Philos. Mag. **86**, 5359 (2006).
[56]	J. N. Connor and R. G. Horn, Farad. Discuss. **123**, 193 (2003).
[57]	R. Manica, J. N. Connor, S. L. Carnie, R. G. Horn, and D. Y. C. Chan, Langmuir **23**, 626 (2007).

Supporting information for
Out-of-contact elastohydrodynamic deformation due to lubrication forces
Yumo Wang[1], Charles Dhong[1], and Joelle Frechette[1,2]
[1]Chemical and Biomolecular Engineering Department, and [2]Hopkins Extreme Materials Institute,
Johns Hopkins University, Baltimore MD 21218.
(Dated November 1, 2015)

## 1. Preparation of PDMS films for the SFA

The PDMS elastomer and curing agent (Dow Corning Sylgard 184 elastomer kit) are mixed with a 10:1 ratio followed by stirring for 5 minutes. The mixture is then placed in vacuum for 20 minutes to remove bubbles, followed by spin-coating on a clean glass coverslip. The PDMS film then is cured at 75ºC for 3 hours and left overnight at room temperature. After curing, the PDM2S is extracted in hexanes for 24 hours to remove the unreacted oligomers.[1] The hexane is then removed via three 15-minutes sonication cycles in 200 proof ethanol. After the sonication, the PDMS is left overnight in a vacuum oven at 75ºC to remove the ethanol. Next, the PDMS is plasma treated for 5s in a home-built plasma reactor at 50W in 0.3 Torr oxygen. This step increases the adhesion between the silver layer on top and the PDMS film. After plasma treatment, the PDMS film is carefully cut from the supporting substrate and glued on the cylindrical disk. a 50 nm silver film is then deposited on the PDMS via thermal vapor deposition. The thickness of the PDMS layer is measured using profilometry prior to the SFA experiments.

Table. S1. Experimental parameters

| Type | Physical parameter | Value |
|---|---|---|
| Fluid (silicone oil) | Viscosity, $\eta_{fluid}$ | 0.2 Pa·s |
| | Density, $\rho$ | 0.98 g/cm$^3$ |
| Thick PDMS film | Thickness, T | 330 μm |
| | Disk 1: $R_1$ | 1.62 cm |
| | Disk 2: $R_2$ | 1.89 cm |
| | $R_h = 2R_1R_2/(R_1+R_2)$ | 1.74 cm |
| | $R_g = R = (R_1R_2)^{1/2}$ | 1.75 cm |
| | Young's modulus, E | 1.08 ± 0.05 MPa |
| | Poisson's ratio, $\nu$ | 0.5 |
| | Viscosity (fitted), $\eta_{PDMS}$ | 0.15 MPa·s |
| Thin PDMS film | Thickness, T | 10.9 μm |
| | Disk 1: $R_1$ | 1.18 cm |
| | Disk 2: $R_2$ | 1.02 cm |
| | $R_h = 2R_1R_2/(R_1+R_2)$ | 1.09 cm |
| | $R_g = R = (R_1R_2)^{1/2}$ | 1.10 cm |
| SFA | Spring constant, k | 165.3 N/m |
| | Drive velocity, V | 69-355 nm/s |
| | Initial separation, $h(0,0)$ | 2.5-3 μm |
| | Maximum motor travel | 8.3 μm |
| | SU-8 thickness | 6-7 um |
| | Mica thickness | 3-10 um |
| | Top silver thickness | 50 nm |

## 2. Multiple-Beam Interferometry (MBI) in the Surface Forces Apparatus (SFA)

We rely on multiple-beam interferometry (MBI) to map the local fluid film thickness within nanometer resolution in the normal direction. In MBI the presence of the two semi-transparent bounding (silver) films facilitates analysis of transmitted white light through the layered system shown in Fig 1B, and due to destructive interferences only certain wavelength are transmitted. The transmitted wavelengths, called fringes of equal chromatic orders (FECO), are a function of the local optical path between the silver layers (thickness and refractive index) which, as illustrated in Fig. S1A, vary with the radial position *r*. Using fast spectral correlation[2] we convert the transmitted wavelengths into the local thickness of the fluid film, *h(r,t)* (Fig. S1A). The silver film on top of the PDMS allows to decouple the deformation of the compliant film from the fluid film thickness. Without the silver this task is challenging because of 1) nearly matching refractive indices between the PDMS and the silicone oil, and 2) solving for both the fluid film thickness and the local PDMS thickness (both within the optical path in the interferometer).

## 3. Contact mechanics of the experimental system.

A. Contact Mechanics experiments to determine the Effective Young's modulus.

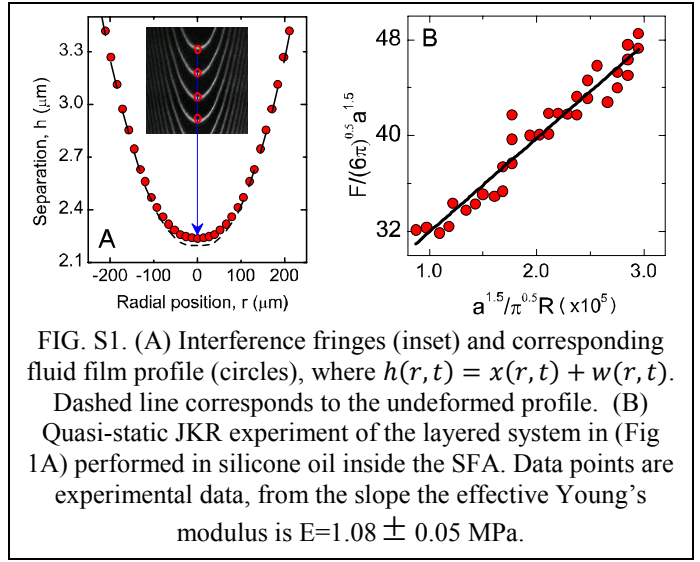

FIG. S1. (A) Interference fringes (inset) and corresponding fluid film profile (circles), where $h(r,t) = x(r,t) + w(r,t)$. Dashed line corresponds to the undeformed profile. (B) Quasi-static JKR experiment of the layered system in (Fig 1A) performed in silicone oil inside the SFA. Data points are experimental data, from the slope the effective Young's modulus is E=1.08 $\pm$ 0.05 MPa.

Prior to the dynamic experiments we perform *in situ* quasi-static JKR contact mechanics experiments[3,4] in the silicone oil with the same interacting surfaces to determine the effective Young's modulus. As shown in Fig. 1B, we obtain an effective Young's modulus of 1.08 ± 0.05 MPa for the PDMS, which is very close to reported values.[5] Attempting to obtain the Young's modulus of compliant supported films from contact mechanics experiments can introduce significant error when the ratio of the contact radius (*a*) is comparable to the film thickness (*T*).[6-8] Here we have at most *a/T*=0.3, and based on predictions we expect this value to overestimate the true value for the Young's modulus by 15-20% [9-11].

B. Effects of the finite thickness of the PDMS film

The ratio of the contact radius to the film thickness is used to estimate the contribution of the underlying rigid substrate on the estimate of the Young's modulus of the PDMS. There are several approximations that have been derived to estimate the apparent increase in the Young's modulus due to the underlying substrate. According to the equation 15 in Ref [9]:

$$f_c(a/T) = (1 + 1.33(a/T) + 1.33(a/T)^3)^{-1} \tag{S1}$$

where $f_c(a/T)$ is the correction factor for compliance of the layered system, here the compliance defined as inverse of $2E^*a$. We inverse the correction factor to the factor for Young's modulus and estimate stiffening of the system to be 21.3%. Another approximation is Eqn 12 in reference [10]:

$$F = \frac{16E}{9} R^{1/2} \delta^{3/2} [1 + 1.133(a/T) + 1.283(a/T)^2 + 0.769(a/T)^3 + 0.0975(a/T)^4], \tag{S2}$$

Where *F* is the force, *E* is the Young's modulus of the PDMS, *R* is the Radius of indenter, and δ is the indentation depth. We then compare this force to a classic semi-infinite form of Hertz contact model and obtain an effective modulus. The average measured contact radius in the JKR experiments is approximately 50 μm, which is to be compared to the 330 μm thickness of the PDMS film. Based on these values, we get an upper bound of *a/T* to ~0.16, therefore the stiffening calculated based on Eqn S2 is 20.14%. This result is very close to the one from Eqn S1.

For a third and final approximation of the stiffening effect, we can take the experimental *R* and *T* into a finite element simulation in ABAQUS 6.13 and perform contact mechanics experiments (See Fig. S2). We mesh the soft compliant film and set the film thickness to be a finite value of 300 μm. We drive the indenter, which is a SU-8 half sphere, quasi-statically and export both the pressure and contacting radius of the two surfaces. Compared to a control experiment which was run on a semi-infinite soft layer with original setting of modulus to be 1MPa and resulting a 0.993 MPa output, the layered system have an output modulus of 1.147 MPa from JKR based on $\frac{F}{\sqrt{6\pi}a^{3/2}} = \frac{4E^*}{3\sqrt{6}}\left(\frac{a^{3/2}}{\sqrt{\pi R}}\right) - \sqrt{\frac{4wE^*}{3}}$, a 14.7% error due to the finite thickness of the PDMS film (Fig. S2). The predictions are in excellent agreement with JKR theory as evidenced by the R-square value that is very close to 1. Combining this simulation with independent calculations based on Refs [9,10], we can say that the Young's modulus value of 1.08+- 0.05 MPa we obtain from experiments includes a stiffening contribution of 15%-20%.

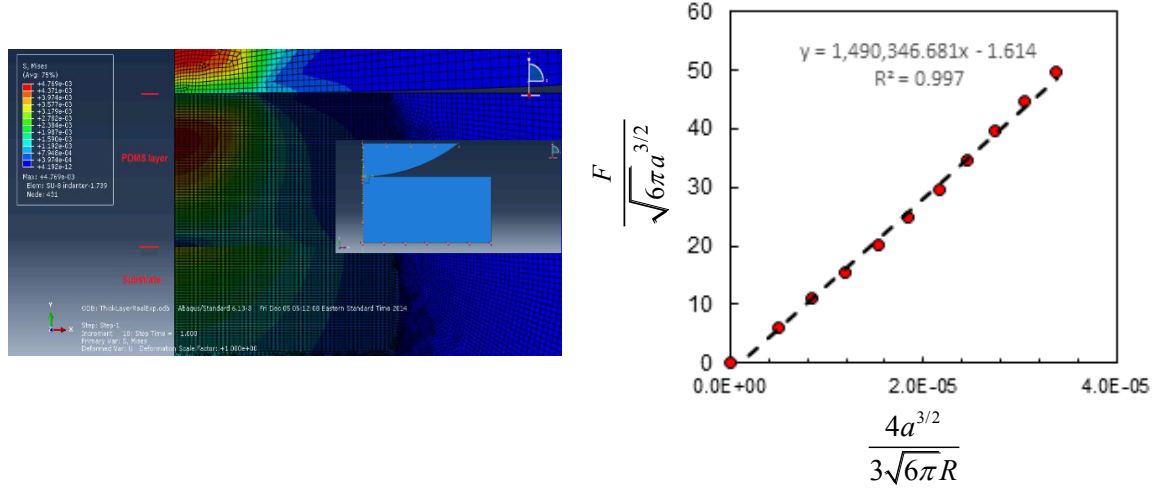

FIG. S2. (Left) Von Mises pressure distribution and schematic of the geometry (inset) of a rigid spherical indenter on a PDMS layer with finite thickness. Color gradient represent the Von Mises stress distribution. The mechanical properties for the materials are 1) Top: Rigid SU-8 indenter: Young's modulus: 2GPa, Poisson ratio: 0.22. Bottom: Soft film on rigid substrate. Thickness: 300 μm. Young's modulus: 1MPa. Poisson ratio: 0.48(typically used Poisson ratio for PDMS is 0.5. Assigning exact value of 0.5 cause problems in our simulation. Here the small change in Poisson ratio does not alter results because the contact radius is small compared to film thickness). The indenter move down stepwise from 0 μm to maximum contacting diameter close to 200 μm. (Right) Output JKR curves from finite elements simulation. From the slope of the curve the effective modulus of the system $E^*$ is obtained, which is the effective modulus of the PDMS if we assume a Poisson ratio to be close to 0.5.

C. Effect of thin silver layer (50nm) on top of the soft layer

We do not expect the 50 nm silver film on the PDMS to affect the load-indentation curves because of the thickness of the silver film compared to that of the PDMS (50 nm vs 330 μm) along with the large radius of curvature in the SFA[11,12]. The silver film has the added advantage of preventing swelling of the PDMS in the oil. It is possible that the top silver layer constrains the deformation of the PDMS layer, and could have an effect on the Young's modulus. To estimate the role played by the silver film, we consider the approximation from Eqn. 11 in Ref [12] :

$$F_f = \pi E_f^* Y \left[ \frac{-6a^6 + 6a^4(R^+)^2 + Y^2 a^4 - 4Y^2 a^2 (R^+)^2}{6((R^+)^2 - a^2)^{5/2}} \right] . \quad (S3)$$

In which, $F_f$ is the effective extra load due to metal layer, $E_f^*$ is the effective modulus of the metal film, $R^+$ is the curvature of the deformation during indentation. Y is the thickness of the metal film on top. Since we have $F_{sys(f)} = F_{(f)} + F_{sys}$ which equals the total load of a system with the additional load due to metal layer and the load due to underlying system. We can then compare the modulus of the PDMS-Silver layers with and without the silver to get:

$$\frac{E^*_{sys(f)}}{E^*_{sys}} = 1 + \frac{3R\pi Y}{4a^3} \cdot \left[ \frac{-6a^6 + 6a^4(R^+)^2 + Y^2 a^4 - 4Y^2 a^2 (R^+)^2}{6((R^+)^2 - a^2)^{5/2}} \right] \quad (S4)$$

Based on Eqn. S4, we estimate the actual importance of silver layer on the apparent Young's modulus. The results show that the influence of the silver layer increases with the contacting area. At the largest area of contact in our experiment (radius ~100 μm), the stiffening due to the silver film is only about 0.4%, we can safely neglect the effect of the silver layer in our application.

D. Effects of the bottom SU-8 surface

Finally, since we are considering a cross-cylinder geometry and treat it as a sphere approaching a wall, the total elastic modulus is determined by both cylinders. In our system (Fig. 2A.), we have a soft and relatively thick top surface ($E$~$1MPa, T = 330\mu m$). To determine the elastic parameter for the whole system, i.e. $\theta = \frac{1-v_{top}^2}{E_{top}} + \frac{1-v_{bottom}^2}{E_{bottom}}$, we find that even when taking the least rigid layer of the bottom surface (Mica, E~2GPa, Poisson ratio 0.22) the effect of bottom layer on θ is still less than 0.1%. Thus, while the bottom surface is a stack of multiple layers, and they might all be affecting each other's to have an effective modulus which may be a composite of all the layers, the influence of these layers on the effective Young's modulus of the system are not important when compared to the PDMS surface. Therefore 1) we don't expect any deformation of the bottom surface, and 2) we do not expect the effective Young's modulus to be influenced by the bottom surface in our experiments.

## 4. Elastohydrodynamics model and numerical algorithm.

For the theoretical description we stay within the lubrication limit as the fluid film thickness is always much smaller than the hydrodynamic radius $\approx \sqrt{2Rh}$. We follow the approach of Davis et al[13] and rely on the fluid pressure distribution, *p(r,t)*, to couple the lubrication equation in axisymmetric radial coordinates (Eqn. S5) with linear elasticity theory on a spherical half-space for axisymmetric pressure distribution (Eqn. S6):

$$\frac{\partial h}{\partial t} = \frac{1}{12\eta r} \frac{\partial}{\partial r}\left[rh^3 \frac{\partial p}{\partial r}\right] \tag{S5}$$

$$w(r) = \frac{4(1-v^2)}{\pi E} \int_0^R \frac{\xi}{r+\xi} K\left[\frac{4r\xi}{(r+\xi)^2}\right] p(\xi) d\xi \tag{S6}$$

where $t$ is the time, $\eta$ is the viscosity of fluid, $v$ is the Poisson's ratio of the elastomer (here $v \approx 0.5$), $E$ is the Young's modulus of the elastomer, $K$ is the elliptical integral of the first kind, and $\xi$ is an integration variable. Here at all times the deformation remains in the small strain limit ($\varepsilon < 0.5\%$) and the radial shear stress on the film is neglected. In the limit of small *w/x* Eqns S5-S6 can be solved analytically by neglecting the contribution of the elastic deformation on the surface profile *h* but a numerical solution is necessary for larger *w/x*.

Lian et al[14] developed an approximate analytical solution by assigning a Hertzian pressure to a freely moving sphere and obtain a solution that is quite close to the numerical solution of Davis[13]. However we cannot use this treatment because our experiments are not performed under constant force, as one of the surface is connected to a motor via a cantilever spring. Therefore, the spring force is balanced by the hydrodynamic force (obtained from the radial integration of the fluid pressure distribution) as shown in Eqn. S7.

$$F = \int_0^R p(r) 2\pi r dr = k\Delta x = k\left(-Vt - h_0 + h - w\right), \tag{S7}$$

Which can be rearranged into:

$$h = \frac{F}{k} + Vt + h_0 + w. \tag{S8}$$

Based on this system of equations (S5-S8) we can calculate the separation at a given time using the initial profile (*h(r)* at *t=0*), the no-slip boundary condition, and known materials properties. To solve numerically the system of equations, we discretize time and lateral positions. At each time increment the values for the prior time step are used as a first guess for the separation and velocity. We then iterate until all $h(r)$ values are within 0.1 nm of the previous iteration, and we use these values to calculate deformation profile, $w(r)$ and the hydrodynamic force, $F(t)$. For the integrations of pressure, we restrict the lateral position to within 10% of total radius $R$ beyond which the normal pressure becomes negligible. We also run convergence tests on both $r$ and $t$ until the maximum error is less than 1%.

There is a logarithmic singularity in Eqn. S6 when $\xi$ comes close to r (the elliptical integral become close to infinity). To address this problem, we follow exactly the approach of Serayssol [15] and generate a nonsingular form for the integration using standard mathematical treatment for these equations. The only difference from Ref. [15] is that $\frac{\pi}{2}\int_0^\infty p(\xi)d\xi$ was calculated numerically here instead of using an approximation. The full derivation is available in Ref. [15] and the salient points are below only for clarity. In equation S6, let:

$$\phi(r,\xi) = \frac{\xi}{r+\xi} K\left[\frac{4r\xi}{(r+\xi)^2}\right], \text{ and} \tag{S9}$$

$$I(r) = \int_0^\infty p(\xi)\phi(r,\xi)d\xi . \tag{S10}$$

Therefore equation S6 become:

$$w(r) = \frac{4(1-v^2)}{\pi E} I(r) . \tag{S11}$$

Rewriting:

$$I(r) = \int_0^\infty [p(\xi)-p(r)]\phi(r,\xi)d\xi + p(r)\int_0^\infty \phi(r,\xi)d\xi , \tag{S12}$$

The expansion of $\phi(r,\xi)$ at $\xi \gg r$ is:

$$\phi(r,\xi) = \frac{\pi}{2} + O\left(\sqrt{\frac{r}{\xi}}\right) , \tag{S13}$$

Thus we can make the form:

$$I(r) = \int_0^\infty [p(\xi)-p(r)]\left[\phi(r,\xi)-\frac{\pi}{2}\right]d\xi + p(r)\int_0^\infty \left[\phi(r,\xi)-\frac{\pi}{2}\right]d\xi + \frac{\pi}{2}\int_0^\infty p(\xi)d\xi \tag{S14}$$

It was shown in Ref. [16], that the integration $\int_0^\infty \left[\phi(r,\xi)-\frac{\pi}{2}\right]d\xi$ is zero, and also that $\frac{\pi}{2}\int_0^\infty p(\xi)d\xi$ can be calculated numerically given a set of pressure values at corresponding $\xi$.

The remaining unsolved part in Eqn. S14 is $\int_0^\infty [p(\xi)-p(r)]\left[\phi(r,\xi)-\frac{\pi}{2}\right]d\xi$, which we call $I_2(r)$, we therefore have:

$$I(r) = \frac{\pi}{2}\int_0^\infty p(\xi)d\xi + I_2(r) . \tag{S15}$$

To solve S15 we need to introduce a large number Y to cut the integration into two parts:

$$I_2(r) = \int_0^Y [p(\xi) - p(r)]\left[\phi(r,\xi) - \frac{\pi}{2}\right] d\xi$$
$$+ \int_Y^\infty [p(\xi) - p(r)]\left[\phi(r,\xi) - \frac{\pi}{2}\right] d\xi \tag{S16}$$

We can further expand the second term and rewrite:

$$I_2(r) = \int_0^Y [p(\xi) - p(r)]\left[\phi(r,\xi) - \frac{\pi}{2}\right] d\xi$$
$$+ \int_Y^\infty p(\xi)\left[\phi(r,\xi) - \frac{\pi}{2}\right] d\xi - p(r)\int_Y^\infty \left[\phi(r,\xi) - \frac{\pi}{2}\right] d\xi \tag{S17}$$

We then integrate the first term numerically. The second term decays as $1/\xi^5$ for small $r$ compared to $\xi$. If we pick $Y > 2$ we can safely neglect this term. The third term in S17 can be integrated analytically by expanding $\phi(r,\xi)$ for large Y:

$$\phi(r,\xi) = \frac{\pi}{2}\left[1 + \frac{1}{4}\frac{r^2}{\xi^2} + O\left(\frac{r^3}{\xi^3}\right)\right]. \tag{S18}$$

Thus:

$$p(r)\int_Y^\infty \left[\phi(r,\xi) - \frac{\pi}{2}\right] d\xi = p(r) \cdot \frac{\pi}{8}\frac{r^2}{Y}, \tag{S19}$$

The final solution for $w(r)$ is then:

$$w(r) = \frac{4(1-\nu^2)}{\pi E}\left[\int_0^Y [p(\xi) - p(r)]\left[\phi(r,\xi) - \frac{\pi}{2}\right] d\xi - p(r) \cdot \frac{\pi}{8} \cdot \frac{r^2}{Y} + \frac{\pi}{2}\int_0^\infty p(\xi) d\xi\right] \tag{S20}$$

Which we used in our algorithm to solve for $w(r)$.

## 5. Viscoelasticity of the PDMS film.

When we treat the elastomer as a viscoelastic solid we modify Eqns. S5-S6 and model the PDMS film's response to an applied load as a spring and dashpot in parallel (Kelvin-Voigt model, see Fig. 1A). We separate the contribution of the fluid pressure applied either to elastic ($p_E(r,t) = E\varepsilon(r,t)$) or viscoelastic deformation $p_V(r,t) = \eta_{PDMS}\, d\varepsilon(t)/dt$, where $\eta_{PDMS}$ is the viscosity and $\varepsilon(r,t)$ is the strain of the PDMS coating. The value for $d\varepsilon(t)/dt$ is approximated using the calculated strain from previous numerical step ($\varepsilon_P$) to get $\frac{\varepsilon - \varepsilon_P}{\Delta t}$ which allows us to determine the fraction of the pressure diverted to the elastic branch, $p_E$. In Eqn. S6, we replace $p$ by the elastic component $p_E$ and continue our calculation. Here $\eta_{PDMS}$ is not known *a priori* and we iterate to find a single $\eta_{PDMS}$ that best describes all the profiles for all drive velocities.

The viscous effect of PDMS on the fluid film thickness is fairly small in our experiments. However we did find that involving viscoelasticity helped obtain better agreement with the experimental results at the beginning of the approach (first few seconds), when the rate of acceleration is highest. This effect, as expected, becomes stronger with increasing drive velocities. See Fig. S3 for details. Acceleration during start up leads to visible viscoelastic contributions, see Fig. S3, where the film deformation lags behind the pressure distribution imposed by the fluid. Because the strain is small during start up, the effect of the viscoelasticity of the PDMS on the fluid film thickness is small. However, a significant fraction of the total pressure gets initially "diverted" into the viscoelastic branch of the Kelvin-Voight model since the ratio of the total pressure applied to elastic deformation ( $p_E/p_{tot}$ ) in the first 5s is less than 0.7.

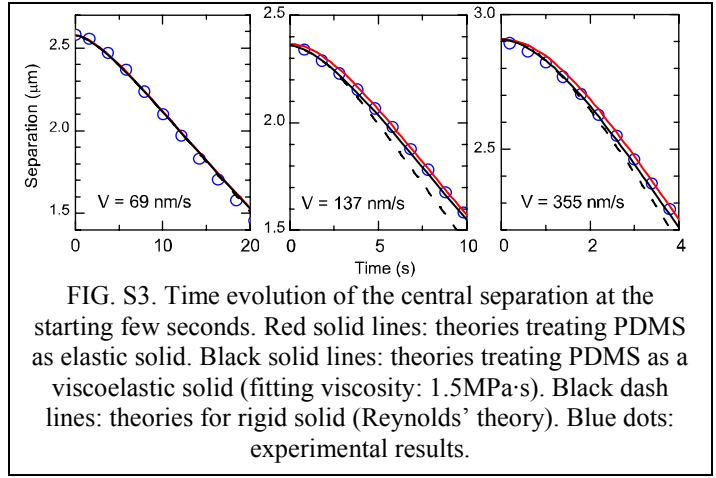

FIG. S3. Time evolution of the central separation at the starting few seconds. Red solid lines: theories treating PDMS as elastic solid. Black solid lines: theories treating PDMS as a viscoelastic solid (fitting viscosity: 1.5MPa·s). Black dash lines: theories for rigid solid (Reynolds' theory). Blue dots: experimental results.

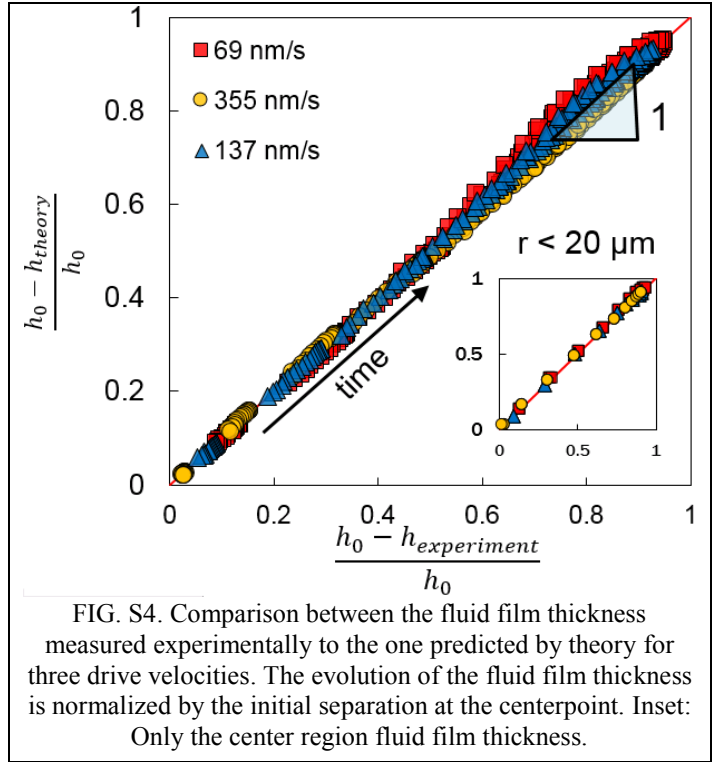

FIG. S4. Comparison between the fluid film thickness measured experimentally to the one predicted by theory for three drive velocities. The evolution of the fluid film thickness is normalized by the initial separation at the centerpoint. Inset: Only the center region fluid film thickness.

## 6. Error analysis

Shown in Fig. S4 are the predicted vs measured change in fluid film thickness, normalized by the initial separation and compiled for all the $h(r,t)$ data (over three drive velocities). The inset of Fig. S4 shows the values obtained only near the centerpoint. The approach to contact in Fig. S4 goes from left to right. We could find that agreement with theory is excellent near the center point (inset). We also see that the deviations are more important close to contact, likely due to the finite thickness of the film and the presence of a rigid underlying substrate. The effect of the underlying substrate on the deformation is clearly visible from the surface profiles near contact (see Fig. S5). For three drive velocities we observe systematic deviations from theory that cannot be accounted for by viscoelastic effects. The shape of the surface profiles show less deformation than expected at the center, which we attribute to the rigidity of substrate, the substrate constrains the surface from becoming as broad as predicted by the theory for half-space.

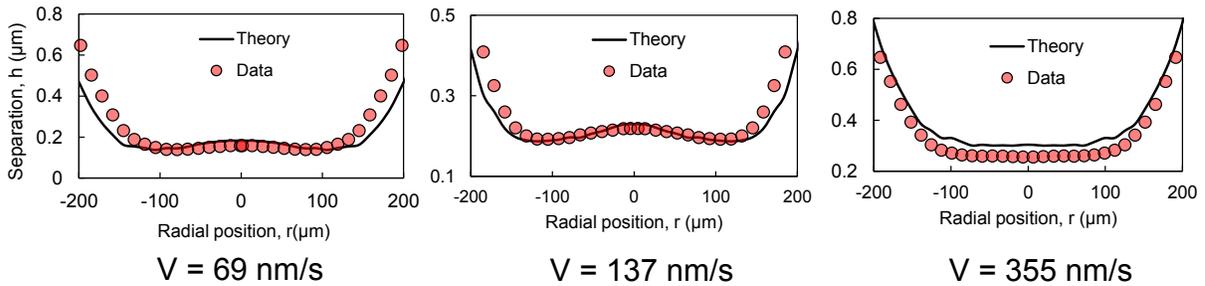

FIG. S5. Individual measured surface profile along with theoretical predictions

The error close to contact cannot be explained by using a different Young's modulus as a fitting parameter. In Fig. S6 we vary the Young's modulus to various values and plot the corresponding surface profiles. Increasing or decreasing the modulus will not improve the fits globally. A larger modulus might lead to better agreement close to contact but would miss the profiles when the surfaces are further away, which is hard to justify (contrast the two panels in Fig. S6). Also the fit in the region of large *r* might be improved with a larger modulus, however it brings significant error in the center region where the pressure is highly concentrated. The fact that that deviation from theory cannot be accounted for by a change in Young's modulus is consistent with contribution of the underlying substrate on the "effective rigidity" that varies with the pressure distribution. To improve the general agreement an effective $E(p)$ would be necessary.

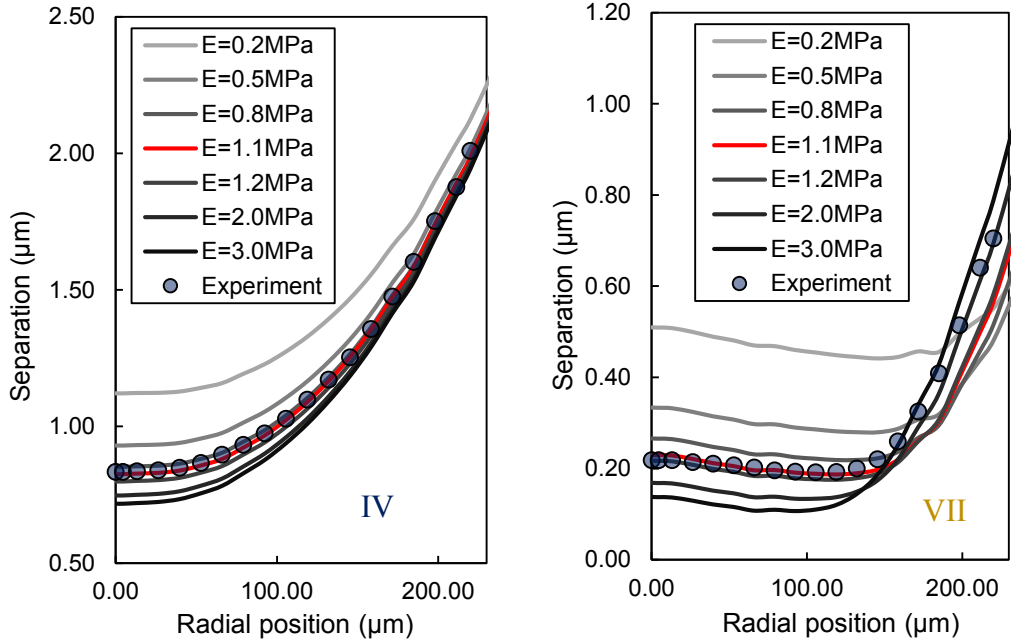

Fig. S6. Theoretical fittings using various modulus. V = 137 nm/s. (Left) t = 18.8s. (Right) t = 53.8s. Roman numerals are the same times as in Fig. 2A.

## 7. Determination of the deformation of the PDMS film from the surface profile.

Shortly after the surfaces start moving, the spatiotemporal deformations are significant and present everywhere in the field of view. It is, therefore, a challenge to estimate $w(r,t)$ from the raw data, $h(r,t)$. We employ two approaches to estimate $w(r,t)$: 1) Assume that $w = 0$ at the limit of our field of view ($r = 200\mu m$), and 2) Calculate $w(r,t)$ from $w(r,t) = h_{exp}(r,t) - x_{th}(r,t)$, here $h_{exp}(r,t)$ is the measured fluid film profile and $x_{th}(r,t)$ is the predicted undeformed position. We can evaluate these two approximations by comparing the measured values with theoretical predictions. Test for the first approximation is shown in Fig. S7 where, following the formalism of Davis et al.[13], the values of the deformation is normalized by the undeformed position, $x(r,t)$ (see Fig. 1A): $w(r,t)/x(r,t)$. We see that assuming that $w(200\mu m, t) = 0$ leads to a systematic error and significantly underestimates the spatiotemporal deformation. Additionally, since the deformation term appear in the force balance equation of SFA, a systematic underestimation of deformation will lead to a systematic overestimation on repulsive force. See Fig. S6, assuming the deformation is zero at the end of our field of view ( $r = 200 \mu m$ ) overestimates the force by 10-20% (open symbols in Fig. S8).

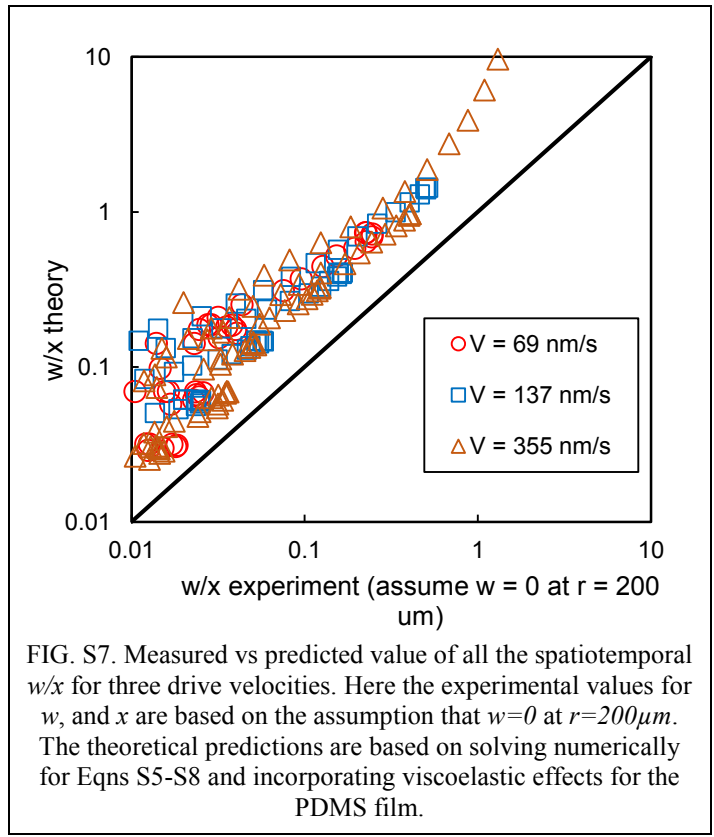

FIG. S7. Measured vs predicted value of all the spatiotemporal $w/x$ for three drive velocities. Here the experimental values for $w$, and $x$ are based on the assumption that $w=0$ at $r=200\mu m$. The theoretical predictions are based on solving numerically for Eqns S5-S8 and incorporating viscoelastic effects for the PDMS film.

Test for the second approximation is shown in Fig. S9, where we compare the experimental values with two versions of the theory: i) the full numerical solution (filled data points), and ii) the analytical approximation proposed by Davis et al [13] (open symbols). The agreement with the full numerical solution in Fig. S9 is excellent. We vary $w/x$ by three orders of magnitude, allowing us to test the validity of the EHD theory in the large deformation limit. As expected the linear approximation fails rather quickly, but interestingly the source of error has two different origin, the first is ignoring the change in the shape of the sphere when predicting the pressure distribution, and the second (and most important one) comes from ignoring the effect of the deformation on the velocity of the surface (compared to the drive velocity). On the other hand, the full numerical solution does well for the whole range of $w/x$, the data in Fig. S9 is compiled from three drive velocities and for all radial positions. We suspect that the main source of error at low $w/x$ comes from limitation in our spatial and lateral resolution. For larger $w/x$ the error comes from the finite thickness of the elastomer layer.

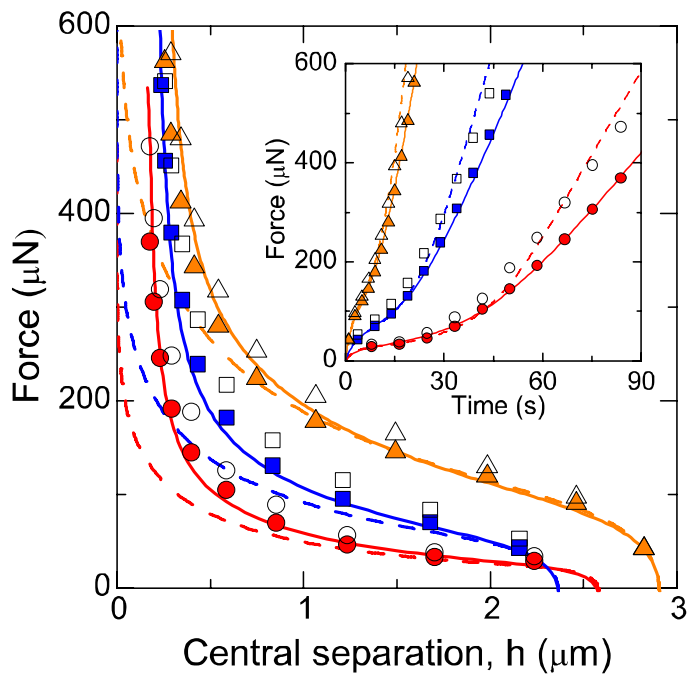

FIG. S8. Error analysis on elastohydrodynamic force between two surfaces as a function of central separation, h. Circles: $V$ = 69 nm/s. squares: $V$ = 137 nm/s. triangles: $V$ = 355 nm/s. The dash lines correspond to Reynolds theory for rigid surfaces and the solid lines represent our predictions based on the full elastohydrodynamic theory. The open symbols are the measured forces assuming that w=0 at r=200μm (Assumption 1).

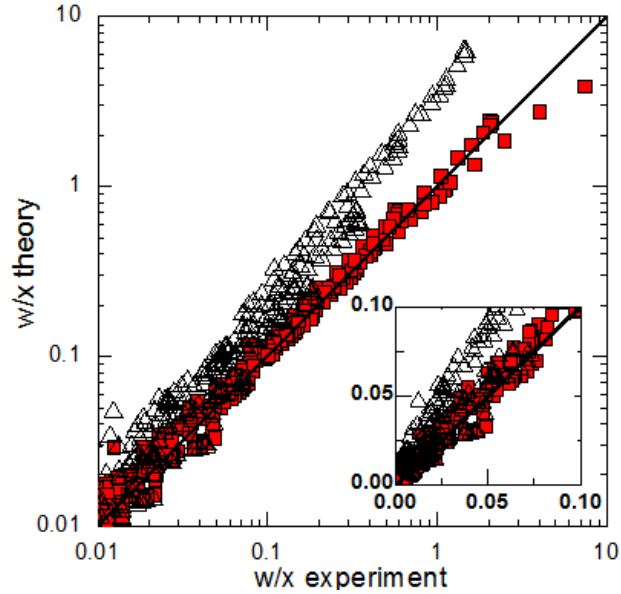

FIG. S9. Comparison between experiment, numerical solution and linear approximation for w/x. Red close squares: comparison of experimental data with full numerical theory. Black open triangles: comparison of experimental data with linearized approximation. Solid black line indicates a slope=1. The inset is an enlargement for small *w/x*.

## 8. References.

[1]     J. N. Lee, C. Park, and G. M. Whitesides, Anal. Chem. **75**, 6544 (2003).
[2]     M. Heuberger, Rev. Sci. Instrum. **72**, 1700 (2001).
[3]     M. K. Chaudhury and G. M. Whitesides, Science **255**, 1230 (1992).
[4]     K. L. Johnson, K. Kendall, and A. D. Roberts, Proc. R. Soc. London, Ser. A **324**, 301 (1971).
[5]     K. Khanafer, A. Duprey, M. Schlicht, and R. Berguer, Biomed. Microdevices **11**, 503 (2009).
[6]     I. Sridhar, K. L. Johnson, and N. A. Fleck, J. Phys. D: Appl. Phys. **30**, 1710 (1997).
[7]     K. L. Johnson and I. Sridhar, J. Phys. D: Appl. Phys. **34**, 683 (2001).
[8]     P. M. McGuiggan, J. S. Wallace, D. T. Smith, I. Sridhar, Z. W. Zheng, and K. L. Johnson, J. Phys. D: Appl. Phys. **40**, 5984 (2007).
[9]     K. R. Shull, Mater. Sci. Eng. R-Rep. **36**, 1 (2002).
[10]    E. K. Dimitriadis, F. Horkay, J. Maresca, B. Kachar, and R. S. Chadwick, Biophys. J. **82**, 2798 (2002).
[11]    G. Barnocky and R. H. Davis, Phys. Fluids **31**, 1324 (1988).
[12]    F. K. Yang, W. Zhang, Y. G. Han, S. Yoffe, Y. C. Cho, and B. X. Zhao, Langmuir **28**, 9562 (2012).
[13]    R. H. Davis, J. M. Serayssol, and E. J. Hinch, J. Fluid Mech. **163**, 479 (1986).
[14]    G. Lian, M. J. Adams, and C. Thornton, J. Fluid Mech. **311**, 141 (1996).
[15]    J.-M. Sérayssol, MS thesis, University of colorado, Boulder (1985)
[16]    B. Hughes and L. White, Q. J. Mech. Appl. Math. **32**, 445 (1979).